\documentclass[conference,a4paper]{IEEEtran}

\usepackage{xcolor}
\usepackage{balance}
\usepackage{microtype}

\usepackage{cite}

\ifCLASSINFOpdf
  \usepackage[pdftex]{graphicx}
\else
  \usepackage[dvips]{graphicx}
\fi
\usepackage{amsmath}
\usepackage{pgfplots}
\pgfplotsset{compat=1.18}
\usepgfplotslibrary{polar}
\definecolor{matlab1}{rgb}{0,0.4470,0.7410}
\definecolor{matlab2}{rgb}{0.8500,0.3250,0.0980}
\definecolor{matlab3}{rgb}{0.9290,0.6940,0.1250}
\definecolor{matlab4}{rgb}{0.4940,0.1840,0.5560}
\definecolor{matlab5}{rgb}{0.4660,0.6740,0.1880}
\definecolor{matlab6}{rgb}{0.3010,0.7450,0.9330}
\definecolor{matlab7}{rgb}{0.6350,0.0780,0.1840}

\usepackage{caption}
\usepackage{subcaption}

\usepackage{tablefootnote}

\usepackage{tabularx}
\usepackage{siunitx}
\usepackage{comment}

\usepackage{tikz}

\newcommand\copyrighttext{%
  \footnotesize \textcopyright \the\year{} IEEE. Personal use of this material is permitted. Permission from IEEE must be obtained for all other uses, including reprinting/republishing this material for advertising or promotional purposes, collecting new collected works for resale or redistribution to servers or lists, or reuse of any copyrighted component of this work in other works.}

\newcommand\copyrightnotice{%
\begin{tikzpicture}[remember picture,overlay]
\node[anchor=south,yshift=10pt] at (current page.south) {\fbox{\parbox{\dimexpr0.75\textwidth-\fboxsep-\fboxrule\relax}{\copyrighttext}}};
\end{tikzpicture}%
}

\begin{document}
\bstctlcite{IEEEexample:BSTcontrol}
\title{3--20\,GHz Wideband Tightly-Coupled Dual-Polarized Vivaldi Antenna Array}

\author{\IEEEauthorblockN{
Niko Lindvall, \textit{Student Member, IEEE},   %
Mikko Heino, \textit{Member, IEEE},   %
and Mikko Valkama, \textit{Fellow, IEEE}    %
}                                     %
\IEEEauthorblockA{%
Tampere Wireless Research Center, Unit of Electrical Engineering, 
Tampere University, Finland}
 \IEEEauthorblockA{niko.lindvall@tuni.fi}
}

\maketitle
\copyrightnotice
\begin{abstract}
Very wideband apertures are needed in positioning, sensing, spectrum monitoring, and modern spread spectrum, e.g., frequency hopping systems.  Vivaldi antennas are one of the prominent choices for the aforementioned systems due to their natural wideband characteristics. Furthermore, tightly-coupled antenna arrays have been researched in the recent years to extend the lower band edge of compact arrays by taking advantage of the strong mutual coupling between the elements especially with dipole elements, but not with dual-polarized Vivaldi antennas. This paper presents a novel tightly-coupled dual-polarized antipodal Vivaldi antenna (TC-AVA) with --6 dB impedance bandwidth of 3 to 20\,GHz. The tight coupling by overlapping the Vivaldi leaves is shown to extend the lower band edge from 3.75 to 3\,GHz and 2.75\,GHz, an improvement of 20\% to 25\% for both polarizations, compared with an isolated antipodal Vivaldi element.

\end{abstract}

\vskip0.5\baselineskip

\section{Introduction}
Radars, sensing, direction finding and positioning are all examples of important applications that benefit from very wide bandwidths, and thus there is a strong interest in improving the frequency range of antenna arrays \cite{10536135,10678888}. For example, \textit{ultra-wideband} (UWB) positioning systems are now common in flagship mobile phones and industrial applications \cite{10622014}. The benefit of UWB systems is increased positioning accuracy, less interference from reflections when calculating the \textit{direction of arrival} (DoA) and increased distance measurement accuracy. Additionally, wideband frequency hopping and spread spectrum systems are popular directions in selected industries, making systems more resilient to interference and jamming \cite{10571810}. Moreover, various spectrum monitoring systems require wide bandwidths for multi-band detection of signals \cite{10709847}.

Typical antenna selections for wideband systems include, for example, wideband dipole elements \cite{elliptic}, Vivaldi antennas\cite{vivaldi}, spiral antennas \cite{spiral, spiral2}, horn antennas \cite{horn} and log-periodic antennas\cite{log-periodic}. Recently, tightly-coupled antenna arrays have been a popular research topic for wideband antenna arrays \cite{10678888,Tightly,10721331}. They utilize the typically undesired coupling between antenna elements to broaden the impedance bandwidth of the antennas. This extends the usable bandwidth of the array towards lower frequencies, compared with individual isolated antenna elements, while still having comparable narrow beamforming capabilities on the upper frequencies. The most researched tightly-coupled arrays are \textit{tightly-coupled dipole arrays} (TCDA), but some tighly-coupled Vivaldi arrays have also been presented.

\begin{figure}[t!]
\centerline{\includegraphics[width=0.45\textwidth]{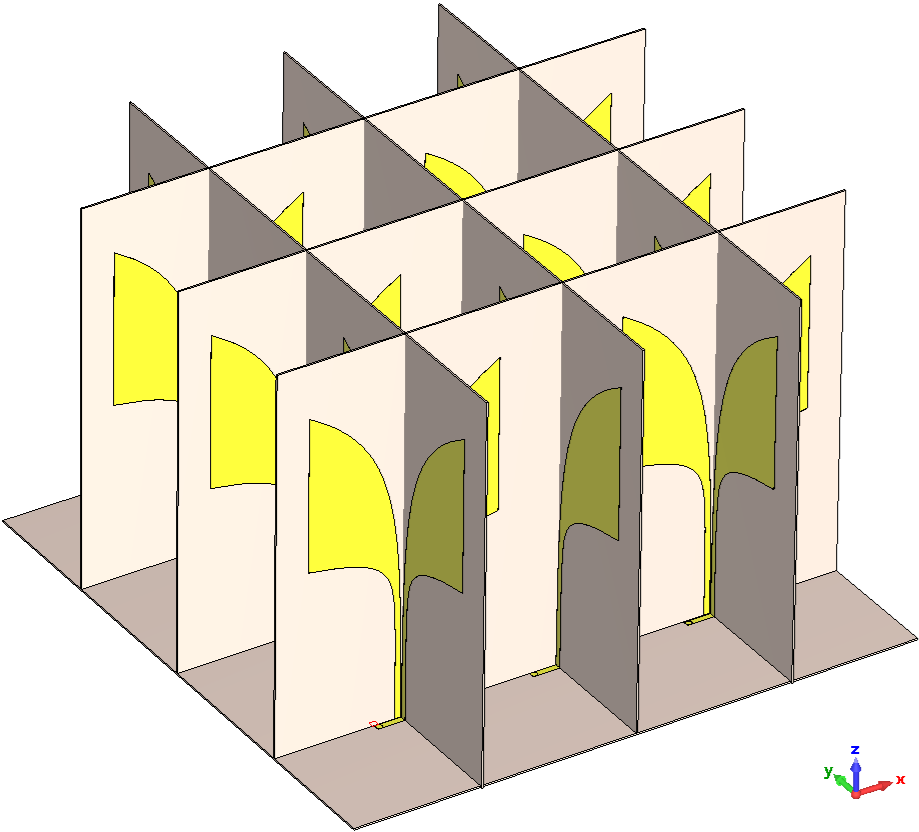}}
\caption{Illustration of the proposed dual-polarized tightly coupled antipodal Vivaldi array concept with overlapping leaves.}
\label{fig:model}
\end{figure}

Typically, dual-polarization is required for modern radar and sensing systems to improve their performance in complex propagation environments. Thus, dual-polarized tightly-coupled dipole arrays have been a popular candidate for spectrum monitoring systems. However, so far \emph{dual-polarized tightly-coupled Vivaldi arrays} have not been designed or presented in the available literature. In traditional arrays, dual-polarization has been implemented for Vivaldi-antennas by using orthogonal PCBs, using single \cite{dual_pol_vivaldi} or dual slots \cite{double_slot_vivaldi}, or 3-dimensional machined full-metal designs \cite{dual_pol_array}, \cite{dual_pol_vivaldi_array2}. \textit{Antipodal Vivaldi-antennas }(AVA) have been used for a single-polarized tightly-coupled array as the design enables easy overlapping of the antenna elements, thus increasing the coupling between elements \cite{Tightly_vivaldi}. However, antipodal Vivaldis are challenging for a dual-polarized array as they do not have an air gap at the center axle which would enable interleaving antennas orthogonally.

This paper seeks to fill this important gap and presents for the first time a tightly-coupled dual-polarized Vivaldi array, conceptually highlighted in Fig.~\ref{fig:model}. The design utilizes perpendicular modified antipodal Vivaldi antennas implemented on PCBs where the feed structure has been modified to enable interleaving for dual-polarization. The adjacent antipodal Vivaldi elements have been implemented on the opposite sides of a PCB to modify the overlap between the antennas and thus the coupling level for tightly-coupled operation. The tight coupling between the elements is shown to extend the lower limit of the --6 dB impedance band from 3.75\,GHz to 3\,GHz, i.e. 25\%, compared to isolated Vivaldi elements, resulting in an operational range for the array from 3\,GHz to 20\,GHz. Special care was needed to design the feeding structure to feed both polarizations from a separate feed-PCB.

\section{Antenna Design}

A classical antipodal Vivaldi antenna \cite{gazit1988improved} was used as a starting point for the design of the antenna element itself. The edge tapering curve and proportions used were adapted from \cite{UWB_Vivaldi} with the inner curve $x_i$ and outer curve $x_o$ defined as
\begin{align}
    x_i &= \pm c_i\mathrm{exp}(k_iy)\mp c_a\\
    x_o &= \pm c_o\mathrm{exp}(k_oy^2)\pm c_b,
\end{align}
with the used parameters described in Table~\ref{tab:dimensions}.

\textls[1]{Dual polarization is a special challenge to implement for antipodal Vivaldis as the opposite sides of the PCB overlap at the feeding point to create a balanced microstrip transmission line, making it difficult to have two concentric orthogonal cross-polarized antennas. The design goal was to create a novel feeding structure that separates the different sides of the feed lines, so that a sufficient gap could be left to place the opposite polarization antenna PCBs concentric with the other polarization. The target was to maximize the impedance bandwidth without a specific target radiation pattern specification. Moreover, the goal was to use a commercially available PCB manufacturing process with widely available materials.}

The first challenge was to separate the feed lines while maintaining good impedance matching at upper frequencies. This required as thin orthogonal PCBs as possible to keep the feed line gap narrow. Thus, 0.254\,mm thick Rogers RO4350B PCB material was selected for the antennas. The input impedance of the symmetric feed of antennas was selected to be 100\,ohms, to get dimensions physically possible to implement, and to enable operation with a standard wideband 2:1 balun, e.g. Mini-Circuits MTX2-183+, with a 50\,ohm unbalanced feed.

\begin{table}[t]
    \centering
    \caption{Dimensions of the prototype design}
    \label{tab:dimensions}
    \begin{tabular}{|c c|c cc|}
        \hline

         & [mm]& & x-pol [mm]& y-pol [mm]\\
        \hline
         $c_i$&$0.03069$&$w_b$&$120.00$&$120.00$\\
         $k_i$&$0.12585$&$w$&$76.39$&$76.39$\\
         $c_a$&$0.48535$&$w_{ol}$&$3.90$&$4.00$\\
         $c_o$&$0.01024$&$w_a$&$28.03$&$28.13$\\
         $k_o$&$0.01058$&$h$&$48.60$&$48.60$\\
         $c_b$&$0.44442$&$h_b$&$56.56$&$56.56$\\
         $w_f$&$0.909$&$w_{sp}$&$0.00$&$0.10$\\
         $l_f$&$4.000$&$h_s$&$0.254$&$0.254$\\
        \hline
    \end{tabular}
\end{table}

\begin{figure}[t]
\centerline{\includegraphics[width=0.5\textwidth]{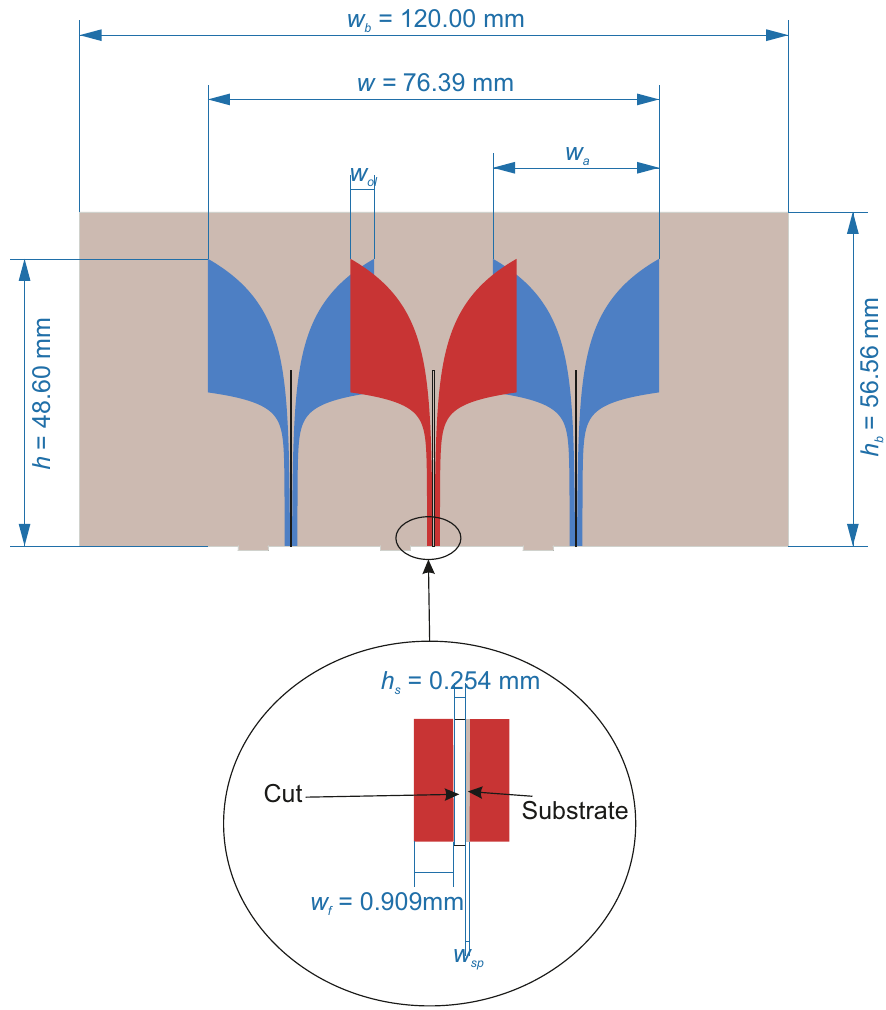}}
\caption{Illustration of the Vivaldi antenna prototype design, with exact dimensions gathered and shown in Table I.}
\label{fig:dimensions}
\end{figure}

\begin{figure}[t]
\centerline{\includegraphics[width=0.35\textwidth]{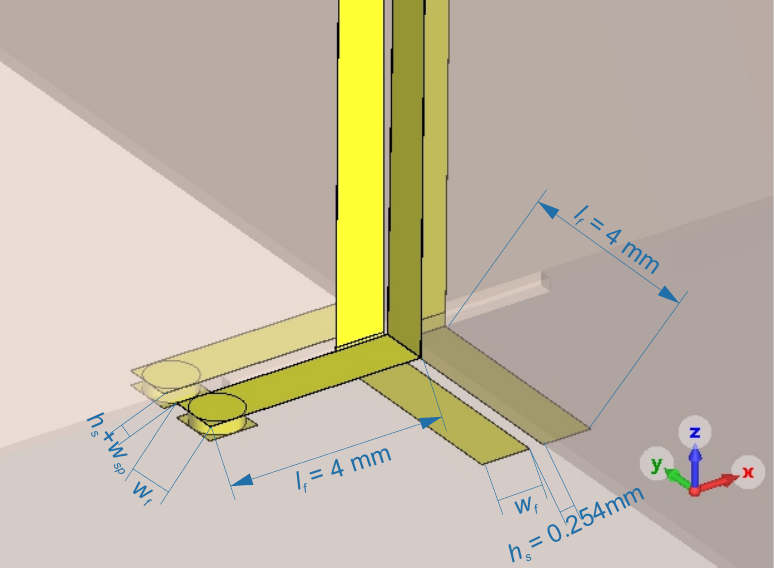}}
\caption{Close-up of the both polarization feeds with their pass-throughs.}
\label{fig:FeedCloseup}
\end{figure}

\begin{figure*}[t!]
    \centering
    \includegraphics{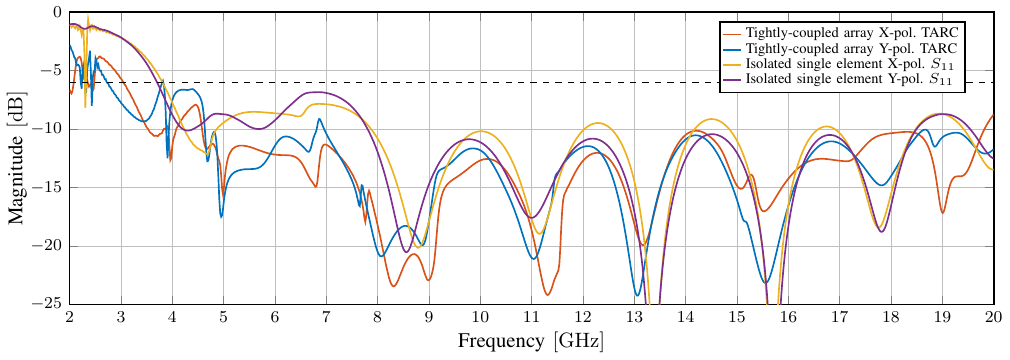}
    \vspace{-2mm}
    \caption{Simulated total antenna reflection coefficients for the proposed dual-polarized tightly coupled array compared with isolated elements.}
    \label{fig:simTARCs}
\end{figure*}

Two concentric AVAs were placed perpendicularly by interleaving  them with the defined 0.254\,mm slots on the PCBs. The design was extended for a 3 $\times$ 3 array to demonstrate the benefit of tight coupling in the array for both polarizations. Adjacent Vivaldis were placed on the opposite sides of the PCB to enable overlap between the flares of Vivaldis and thus to enable control of the coupling level. The overlap and the Vivaldi dimensions were optimized together to obtain the widest impedance bandwidth possible. Figure \ref{fig:model} shows the simulation model with 3 $\times$ 3 array elements for dual-polarized operation. The optimized design parameters of the antenna array are shown in Fig. \ref{fig:dimensions} with the values of the parameters in Table \ref{tab:dimensions}. The overlap of elements $w_{ol}$ was selected as 4\,mm for y-polarization based on the optimization of the bandwidth with simultaneous excitation of the antenna elements. It was decided to keep the same antenna spacing for both polarizations, while the element overlap $w_{ol}$ was reduced to 3.9\,mm for the x-polarization due to the additional spacing required to accommodate the microstrip line required for feeding the other polarization.

\section{Feed design}

After designing the concentric Vivaldis, a major challenge was to design the feed structure in such a way that feeding both polarizations was practically possible without the feed lines crossing. The feeding structure was designed to be implemented on a third perpendicular PCB to which the two antenna PCBs attach, with separate passthroughs for both polarizations keeping the balanced lines intact.

The detailed feed structure is presented in Fig. \ref{fig:FeedCloseup} with additional dimensions presented in Table \ref{tab:dimensions}. For x-polarized antenna elements, the feed lines go through a slot to the other side of the feed PCB where they connect to a balanced strip line on the bottom side forming a 90-degree joint. For y-polarized antenna elements, the antenna feed lines form a 90-degree joint with a balanced strip line on the top side of the feed PCB, with the cross-polarized PCB placed between the lines. The feed line is then transitioned to the bottom side of the feed PCB with PCB vias, 
continuing in a balanced line separated from the other polarization.

\section{Results and Discussion}
\subsection{Results}
In simulations, the feed port is connected to the balanced microstrip line shown in Figure \ref{fig:FeedCloseup} on the feed PCB. Figure \ref{fig:simTARCs} shows the \textit{total active reflection coefficient} (TARC) \cite{tarc} calculated from the simulated S-parameters based on \cite{tarc_correct}, for both polarizations of the tightly coupled AVA array separately when equi-phase feeding 3 elements in the centre row. As comparison, the reflection coefficients for individual isolated AVA elements for both polarizations are shown. For the array design, the --6 dB impedance bandwidth extends from 3 to 20\,GHz, for both polarizations and further down to 2.74\,GHz for y-polarization.

\textls[-1]{The --10\,dB impedance bandwidth extends from 4.6\,GHz to 19.8\,GHz for x-polarization, and from 4.6\,GHz to 20\,GHz for y-polarization, with an exception between 6.81 to 6.95\,GHz where it drops to --9\,dB. For both polarizations, --6\,dB bandwidth of the simulated TARC can extend over 30\,GHz, however, the feeding structure such as a balun will limit the upper frequency range.}

As seen in Fig. \ref{fig:simTARCs}, the --6\,dB bandwidth of a single isolated element ranges from 3.75 to 20\,GHz. Thus, the tight coupling extends the lower limit of the --6\,dB impedance band up to 0.75\,GHz or 20\% for the x-polarization and up to 1\,GHz or 25\% for y-polarization.

\begin{figure}
    \centering
    \includegraphics{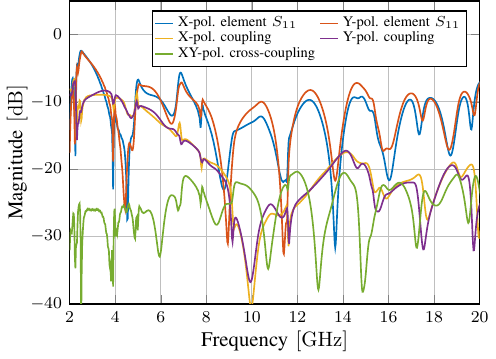}
    \caption{Simulated reflection coefficients, and coupling levels for both polarization elements in the array.}
    \label{fig:arrayelementsparams}
\end{figure}

Figure \ref{fig:arrayelementsparams}, in turn, shows the S-parameters for an element in the middle of the array for both polarizations. The reflection coefficients are improved in the lower band compared with the isolated elements in Fig. \ref{fig:simTARCs}. The coupling is high at --8\,dB between adjacent elements at the low band edge, which is used to improve the reflection coefficient within the array and the total active reflection coefficient. The coupling level is mostly around the --20\,dB level on the upper band, with frequencies higher than 9\,GHz, to enable feeding the elements separately at higher frequencies.  Cross-coupling level between the polarizations remains below --20\,dB.

\begin{figure}
    \centering
    \includegraphics{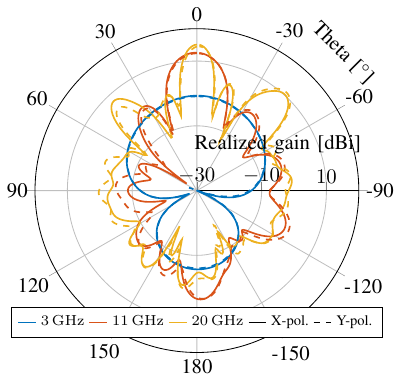}
    \caption{The simulated realized gain radiation patterns of the array in E-plane.}
    \label{fig:patterns}
\end{figure}

\begin{figure}
    \centering
    \includegraphics{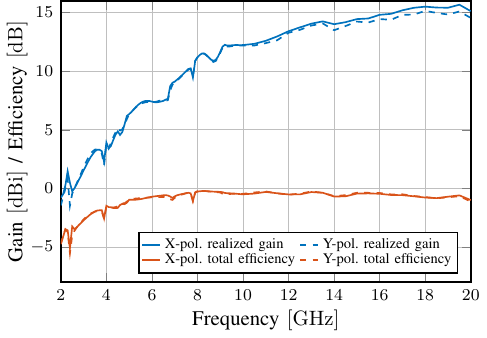}
    \caption{The simulated maximum realized gain and total efficiency of the prototype across frequency.}
    \label{fig:efficiency}
\end{figure}

Figure \ref{fig:patterns} shows the realized gain radiation pattern of the array at the operating band edges and in the middle of the band with equi-phase feeding. At the lower band edge of 3\,GHz, the element spacing of 24\,mm corresponds to 0.24$\lambda_0$ in contrast to 1.6$\lambda_0$ at 20\,GHz. The element spacing of 24,13mm corresponds to half wavelength at 6.2\,GHz. With higher frequencies, sidelobes start to appear as seen in the Figure. Figure \ref{fig:efficiency} shows the total efficiency and the maximum realized gain of the array across the whole operational band. For both polarizations, the total efficiency is better than --1\,dB from 5 GHz up, and gradually goes down to --2.5\,dB at 3\,GHz.

Finally, Table \ref{tab:conparison} shows a comparison of the key performance figures with reference papers. The arrays are of different sizes both in number of elements and in aperture size. To scale the shown gain values, the realized gain is calculated using the VSWR or reflection coefficient from the papers where the realized gain is explicitly not shown. As the array factor effects the gain performance of different size arrays, the realized gain is scaled by a theoretically calculated broadside array factor of $AF = 2Nd/\lambda$ to make the gain figures comparable. Also, some arrays include passive elements that widen the effective aperture at the lower frequencies due to the mutual coupling, which is not compensated. The bandwidth achieved in this work is comparable to the previously presented single-polarized TCVA \cite{Tightly_vivaldi} and the conventional dual-polarized Vivaldi arrays \cite{dual_pol_vivaldi} and \cite{dual_pol_vivaldi_array2} with a higher scaled realized gain in the low band. The proposed antenna has a wider bandwidth than the recently presented individual Vivaldi antenna elements. 

\subsection{Towards Physical Prototype Implementation}

\begin{figure*}[h]
     \centering
     \begin{subfigure}[b]{0.45\textwidth}
         \centering
         \includegraphics[width=\textwidth]{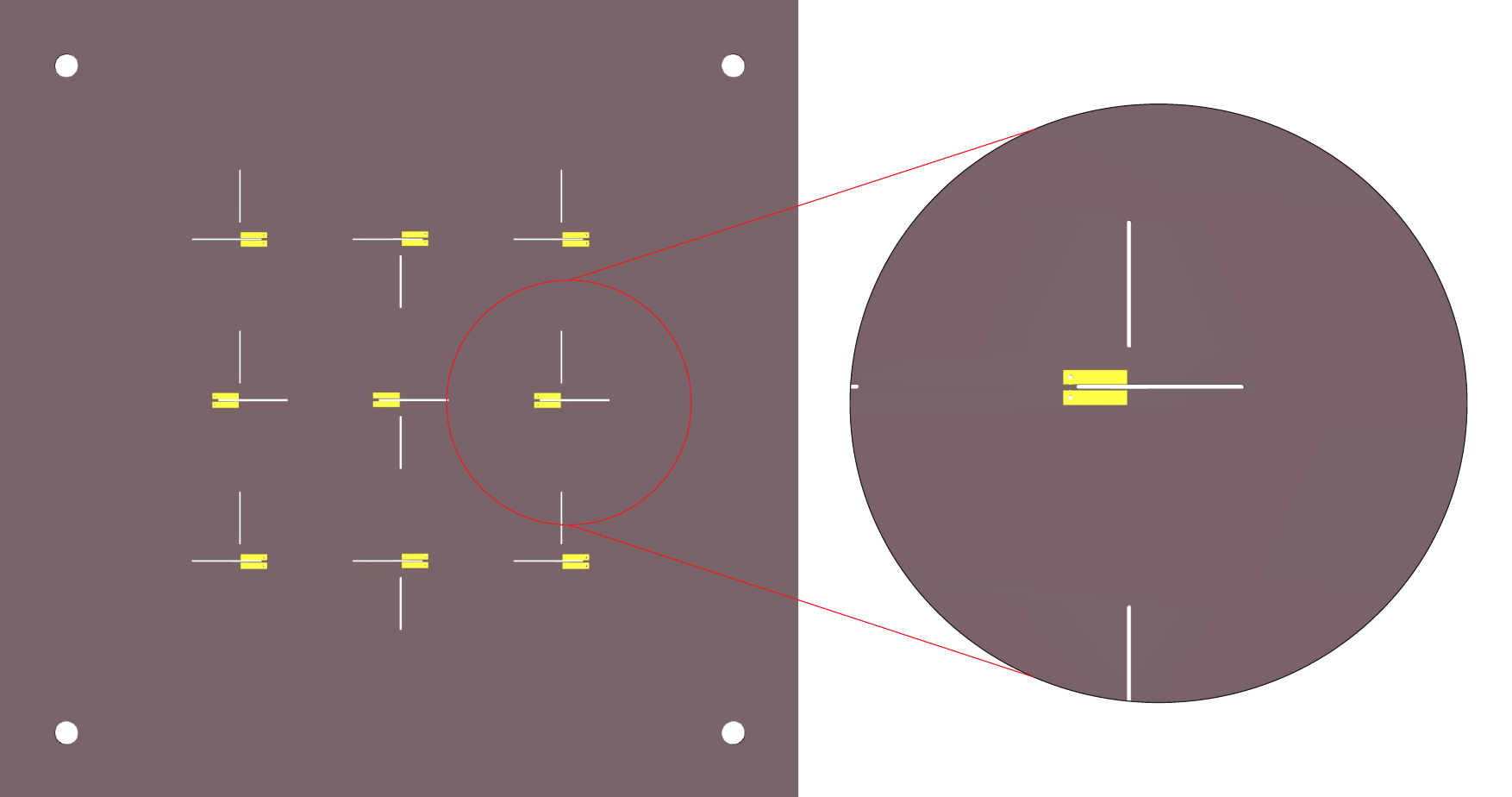}
         \caption{Top}
         \label{fig:Top}
     \end{subfigure}
     \begin{subfigure}[b]{0.45\textwidth}
         \centering
         \includegraphics[width=\textwidth]{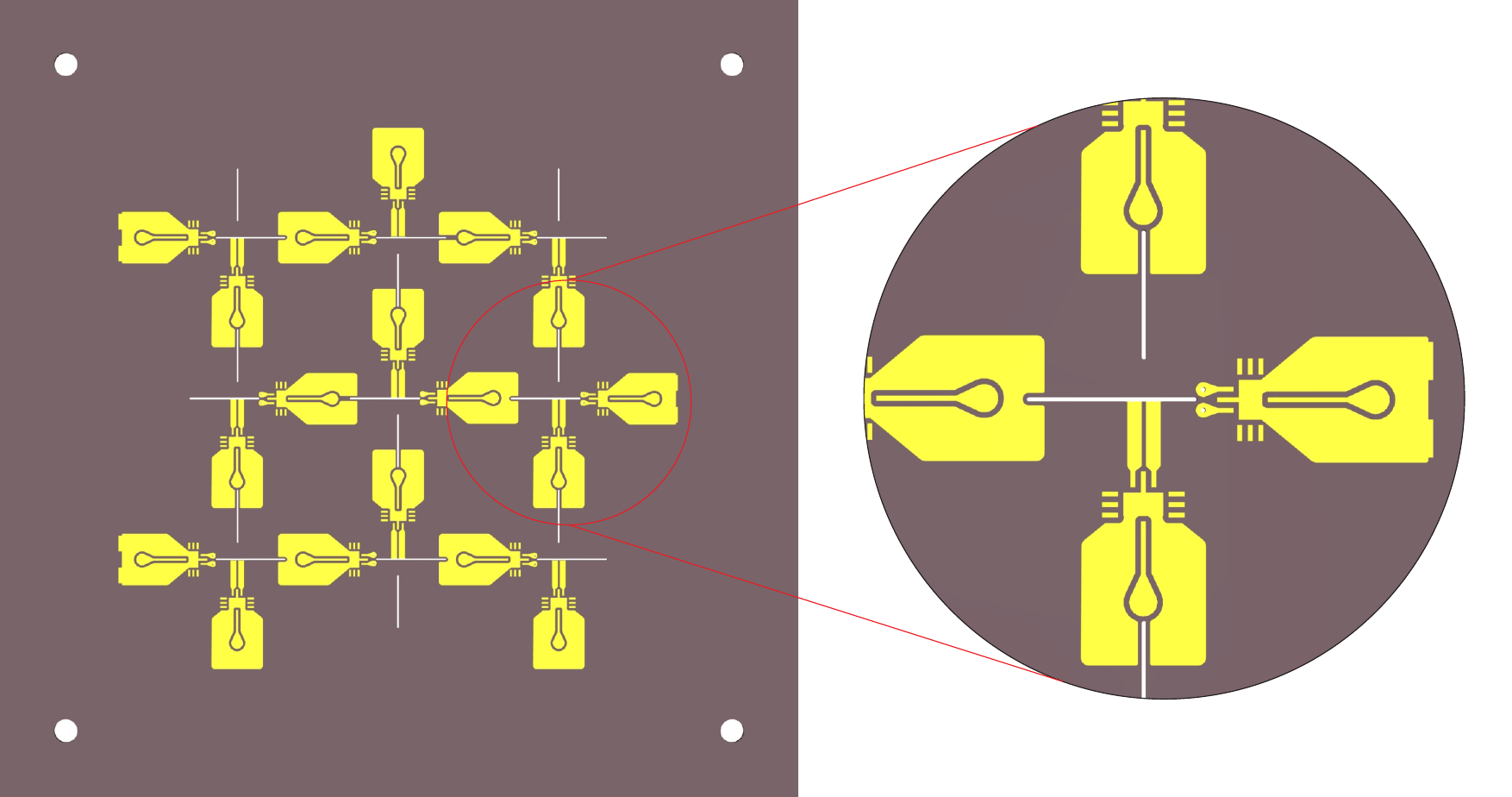}
         \caption{Bottom}
         \label{fig:Bottom}
     \end{subfigure}
        \caption{The feed PCB for the array with separate balun and SMA connector footprint for both polarizations of each element.}
        \label{fig:feedpcb}
\end{figure*}

In our upcoming future work, we will pursue physical prototype implementation and corresponding measurements of the developed tightly-coupled dual-polarized Vivaldi array. Figure \ref{fig:dimensions} shows already the designed realistic antenna PCB for y-polarization featuring slots between the Vivaldi leaves, enabling perpendicular interleaving with the other polarization PCB, which has the slots on the opposite side. Figure \ref{fig:feedpcb} shows both sides of the realistic feed PCB designed for the prototype with mounting slots designed to accommodate the overhangs at the bottom of the antenna PCBs in Figure \ref{fig:dimensions}. 

The whole feeding PCB is designed on a similar 0.254~mm thick 120 mm $\times$ 120 mm sized Rogers 4350B PCB to eliminate the need to change the dimensions of the microstrip line compared to the antenna PCBs. The balanced strip lines on the feed PCB are soldered with 90-degree joints with the antenna PCB feed lines. At the bottom of the feed PCB, the balanced 100\,$\Omega$ lines connect to a 2:1 wideband balun chip, e.g. Mini-Circuits MTX2-183+, enabling the connection to a 50\,$\Omega$ surface-mount SMA coaxial connector.

After combining the 7 PCBs, the complete 3$\times$3 array assembled would resemble the model in Fig. \ref{fig:model} with separate SMA feeds for both polarizations on the bottom side. The modular design enables cost-effective manufacturing with standard PCB processes. Completing and measuring an actual physical prototype is the main topic in our ongoing research.

\section{Conclusion}
This paper presented a novel tightly-coupled antipodal Vivaldi antenna array with an extended bandwidth of 3 to 20\,GHz, with a 20-25\% lower band edge improvement using mutual coupling obtained for both polarizations, compared to an individual antenna element.
The dual-polarization of such a tightly-coupled antenna array has not been implemented previously in the literature, and especially the feeding structure design proved to be challenging. In the proposed design, several perpendicular PCBs are connected together with slots and via passthroughs to enable separate feeding of both antennas from the bottom of the feed PCB. An actual physical prototype implementation, together with the corresponding measurements and their comparisons against simulated results from an important topic for our future work.

\begin{table*}[t]
    \begin{center}
    \caption{Comparison of relevant prior work}
    \label{tab:conparison}
    \begin{tabular}{|c| c S[table-format=2.1] >{\raggedleft}p{1cm} @{~$\sim$~} p{1.2cm} >{\raggedleft}p{1.5cm} >{\raggedleft}p{1cm} r @{\hspace{0.5cm}}|}
        \hline

        Antenna array type & Elements & {~~~~Element separation} & \multicolumn{2}{p{3cm}}{\centering Array BW}        & \multicolumn{3}{c|}{Realized gain divided by array factor}\\
                           &          &                          & \multicolumn{2}{c}{($S_{11} <= -10\,\mathrm{dB}$)}   & low $f$ & \makebox[0.9cm][l]{middle $f$} & high $f$\\
        \hline
         \textbf{Proposed dual-pol. TCVA} & $3\times 3$ & 24.1\,mm & 3 & 20\,GHz & --0.1\,dBi & 5.4\,dBi & 5.3\,dBi\\
         Single-pol. TCVA \cite{Tightly_vivaldi} & $6\times 1$ & 95.0\,mm & ~0.51 & 18\,GHz & --10\,dBi\makebox[0pt][l]{$^*$} & 4.0\,dBi\makebox[0pt][l]{$^*$} & 0.0\,dBi\makebox[0pt][l]{$^*$}\\
         Dual-pol. TCDA \cite{Tightly} & $ 8 \times 8$\makebox[0pt][l]{ $^\dagger$}& 14.0\,mm & 2 & 18\,GHz & --0.7\,dBi & 8.3\,dBi & 5.2\,dBi\\
         Full-metal dual-pol. Vivaldi array \cite{dual_pol_array} & $ 8 \times 8$ & 18.2\,mm & 0.5 & 9\,GHz & --9.9\,dBi\makebox[0pt][l]{$^*$}& 6.9\,dBi\makebox[0pt][l]{$^*$}&N/A~~\\
         Full-metal dual-pol. Vivaldi array \cite{dual_pol_vivaldi_array2} & $ 8 \times 8$\makebox[0pt][l]{ $^\ddagger$} & 12.3\,mm & 3 & 18\,GHz &  --3.2\,dBi& 2.6\,dBi & 6.1\,dBi\\
         Single dual-pol. Vivaldi \cite{dual_pol_vivaldi} & $ 1 \times 1$ & 220.0\,mm & 0.5 & 7.5\,GHz & 4.0\,dBi & 10\,dBi & N/A~~\\
         Single element Vivaldi\cite{UWB_Vivaldi} & $ 1 \times 1$  & 35.5\,mm & 3.1 & 12\,GHz & --2.3\,dBi & 5.4\,dBi & N/A~~\\
         Double slot single-pol. Vivaldi \cite{double_slot_vivaldi} & $ 1 \times 1$ & 42.5\,mm & 5.4 & 16.4\,GHz & N/A~~~ & 10\,dBi  & 11\,dBi\\

        \hline
    \end{tabular}

    \footnotesize{$^*$ Single element in an array, $^\dagger$ Total elements 2664, including passive elements, $^\ddagger$ Total elements $10\times 10$, including passive elements}\\
    \end{center}
\end{table*}

\bibliographystyle{IEEEtran}
\bibliography{eucap2025}

\end{document}